\documentclass[runningheads,fleqn]{svmult}

\usepackage{makeidx}   
\usepackage{graphicx}  
\usepackage{subeqnar}  
\usepackage{multicol}  
\usepackage{physproc}  
\makeindex             

\begin{document}
\title*{Ensemble optimization techniques for the simulation of slowly equilibrating systems}
\toctitle{Ensemble optimization techniques\protect\newline for the simulation of slowly equilibrating systems}
%
%
\titlerunning{Ensemble optimization techniques}
%
\author{S. Trebst\inst{1} 
\and D. A. Huse\inst{2}
\and E. Gull\inst{3}
\and H. G. Katzgraber\inst{3}
\and U. H. E. Hansmann\inst{4,5}
\and M. Troyer\inst{3}}
\authorrunning{S. Trebst et al.}
%
%
\institute{Microsoft Research and Kavli Institute for Theoretical Physics, \\
University of California, Santa Barbara, CA 93106, USA
\and Department of Physics, Princeton University, Princeton, NJ 08544, USA
\and Theoretische Physik, Eidgen\"ossische Technische Hochschule Z\"urich, \\
CH-8093 Z\"urich, Switzerland
\and Department of Physics, Michigan Technological University, Houghton, \\
MI 49931, USA
\and John-von-Neumann Institute for Computing, Forschungszentrum J\"ulich, D-52425 J\"ulich, Germany}

\maketitle              

\begin{abstract}
Competing phases or interactions in complex many-particle systems can result in free energy barriers that strongly suppress thermal equilibration. Here we discuss how extended ensemble Monte Carlo simulations can be used to study the equilibrium behavior of such systems. Special focus will be given to a recently developed adaptive Monte Carlo technique that is capable to explore and overcome the entropic barriers which cause the slow-down. We discuss this technique in the context of broad-histogram  Monte Carlo algorithms as well as its application to replica-exchange methods such as parallel tempering.
We briefly discuss a number of examples including  low-temperature states of magnetic systems with competing interactions and dense liquids. 
\end{abstract}

\section{Introduction}

The free energy landscapes of complex many-body systems with competing phases or interactions are often characterized by many local minima that are separated by entropic barriers. The simulation of such systems with conventional Monte Carlo \cite{MonteCarlo} or molecular dynamics \cite{MolecularSimulation} methods is slowed down by long relaxation times due to the suppressed tunneling through these barriers.
While at second order phase transitions this slow-down can be overcome by improved updating techniques, such as cluster updates \cite{SwendsenWang,Wolff}, this is not the case for systems which undergo a first-order phase transition or for systems that exhibit frustration or disorder.
For these systems one instead aims at improving the way that relatively simple, local updates are accepted or rejected in the sampling process by introducing artificial statistical ensembles such that tunneling times through barriers are reduced and autocorrelation effects minimized.
In the following we discuss recently developed techniques to find statistical ensembles that optimize the performance of Monte Carlo sampling, first in the context of broad-histogram Monte Carlo algorithms and then outline how these methods can be applied in the context of replica-exchange or parallel-tempering algorithms.

\section{Extended Ensemble Methods}
\label{sec:extended}

Let us consider a first-order phase transition, such as in a two-dimensional $q$-state Potts model \cite{Wu} with a Hamiltonian
\begin{equation}
H=-J\sum_{\langle i,j\rangle}\delta_{\sigma_i\sigma_j},
\end{equation}
where the spins $\sigma_i$ can take the integer values $1,\ldots,q$. For $q>4$ this model exhibits a first-order phase transition, accompanied by exponential slowing down of conventional local-update algorithms. The exponential slow-down is caused by the free-energy barrier between the two coexisting meta-stable states at the first-order phase transition. 

This barrier can be quantified by considering the energy histogram
\begin{equation}
H_{\rm canonical}(E) \propto g(E)P_{\rm Boltzmann}(E) = g(E)\exp(-\beta E) \;,
\end{equation}
which is the probability of encountering a configuration with energy $E$ during the Monte Carlo simulation. The density of states is given by
\begin{equation}
g(E)=\sum_c\delta_{E,E(c)} \;,
\end{equation}
where the sum runs over all configurations $c$. Away from first-order phase transitions, $H_{\rm canonical}(E)$ has approximately Gaussian shape, centered around the mean energy. At first-order phase transitions, where the energy jumps discontinuously, the histogram $H_{\rm canonical}(E)$ develops a double-peak structure. The minimum of $H_{\rm canonical}(E)$ between these two peaks, which the simulation has to cross in order to go from one phase to the other, becomes exponentially small upon increasing the system size. This leads to exponentially large tunneling and autocorrelation times.

This tunneling problem at first-order phase transitions can be aleviated by extended ensemble techniques which aim at broadening the sampled energy space. Instead of weighting a configuration $c$ with energy $E=E(c)$ using the Boltzmann weight  $P_{\rm Boltzmann}(E) = \exp(-\beta E)$ more general weights $P_{\rm extended}(E)$ are introduced which define the extended ensemble. The configuration space is explored by generating a Markov chain of configurations 
\begin{equation}
  c_1 \rightarrow c_2 \rightarrow \ldots \rightarrow c_i \rightarrow c_{i+1} \rightarrow \ldots \;, 
  \label{eq:crw}
\end{equation}
where a move from configuration $c_1$ to $c_2$ is accepted with probability 
\begin{equation}
  P_{\rm acc}(c_1 \rightarrow c_2) = \min \left[ 1,\frac{P(c_2)}{P(c_1)} \right] = \min \left[ 1,\frac{W_{\rm extended}(E_2)}{W_{\rm extended}(E_1)} \right] \;.
\end{equation} 
In general, the extended weights are defined in a single coordinate, such as the energy, thereby projecting the random walk in configuration space to a random walk in energy space
\begin{equation}
  E_1=E(c_1) \rightarrow E_2 \rightarrow \ldots \rightarrow E_i \rightarrow E_{i+1} \rightarrow \ldots \;. 
  \label{eq:erw}
\end{equation}
For this random walk in energy space a histogram can be recorded which has the characteristic form  
\begin{equation}
H_{\rm extended}(E) \propto g(E) W_{\rm extended}(E) \;,
\end{equation}
where the density of states $g(E)$ is fixed for the simulated system. 

One choice of generalized weights is the multicanonical ensemble \cite{MultiCanonical} where the weight of a configuration $c$ is defined as $W_{\rm multicanonical}(c)\propto1/g(E(c))$. The multicanonical ensemble then leads to a flat histogram in energy space
\begin{equation}
H_{\rm multicanonical}(E) \propto g(E)W_{\rm multicanonical}(E) = g(E){1\over g(E)}={\rm const.}
\end{equation}
removing the exponentially small minimum in the canonical distribution. After performing a simulation, measurements in the multicanonical ensemble are reweighted by a factor $W_{\rm Boltzmann}(E)/W_{\rm multicanonical}(E)$ to obtain averages in the canonical ensemble.

Since the density of states and thus the multicanonical weights are not known initially, a scalable algorithm to estimate these quantities is needed. The Wang-Landau algorithm \cite{WangLandau} is a simple but efficient iterative method to obtain good approximates of the density of states $g(E)$ and the multicanonical weights $W_{\rm multicanonical}(E)\propto1/g(E)$. 
Besides overcoming the exponentially suppressed tunneling problem at first-order phase transitions, the Wang-Landau algorithm calculates the generalized density of states $g(E)$ in an iterative procedure. The knowledge of the density of states $g(E)$ then allows the direct calculation of the free energy from the partition function, $Z=\sum_Eg(E) e^{-\beta E}$. The internal energy, entropy, specific heat and other thermal properties are easily obtained as well, by differentiating the free energy. 
By additionally measuring the averages $A(E)$ of other observables $A$ as a function of the energy $E$, thermal expectation values can be obtained at arbitrary inverse temperatures $\beta$ by performing just a single simulation:
\begin{equation}
\langle A(\beta) \rangle = \frac{\sum_EA(E)g(E) e^{-\beta E}}{\sum_Eg(E) e^{-\beta E}} \;.
\label{eq:anyt}
\end{equation}

\section{Markov chains and random walks in energy space}

The multicanonical ensemble and Wang-Landau algorithm both project a random walk in high-dimensional configuration space to a one-dimensional random walk in energy space where all energy levels are sampled equally often. It is important to note that the random walk in configuration space, equation~(\ref{eq:crw}), is a biased Markovian random walk, while the projected random walk in energy space, equation~(\ref{eq:erw}), is non-Markovian, as memory is stored in the configuration. This becomes evident as the system approaches a phase transition in the random walk: While the energy no longer reflects from which side the phase transition is approached, the current configuration may still reflect the actual phase the system has visited most recently. In the case of the ferromagnetic Ising model, the order parameter for a given configuration at the critical energy $E_c \sim -1.41 N$ (in two space dimensions) will reveal whether the system is approaching the transition from the magnetically ordered (lower energies) or disordered side (higher energies).

\begin{figure}[t]
  \begin{center}
    \includegraphics[width= 0.75\textwidth]{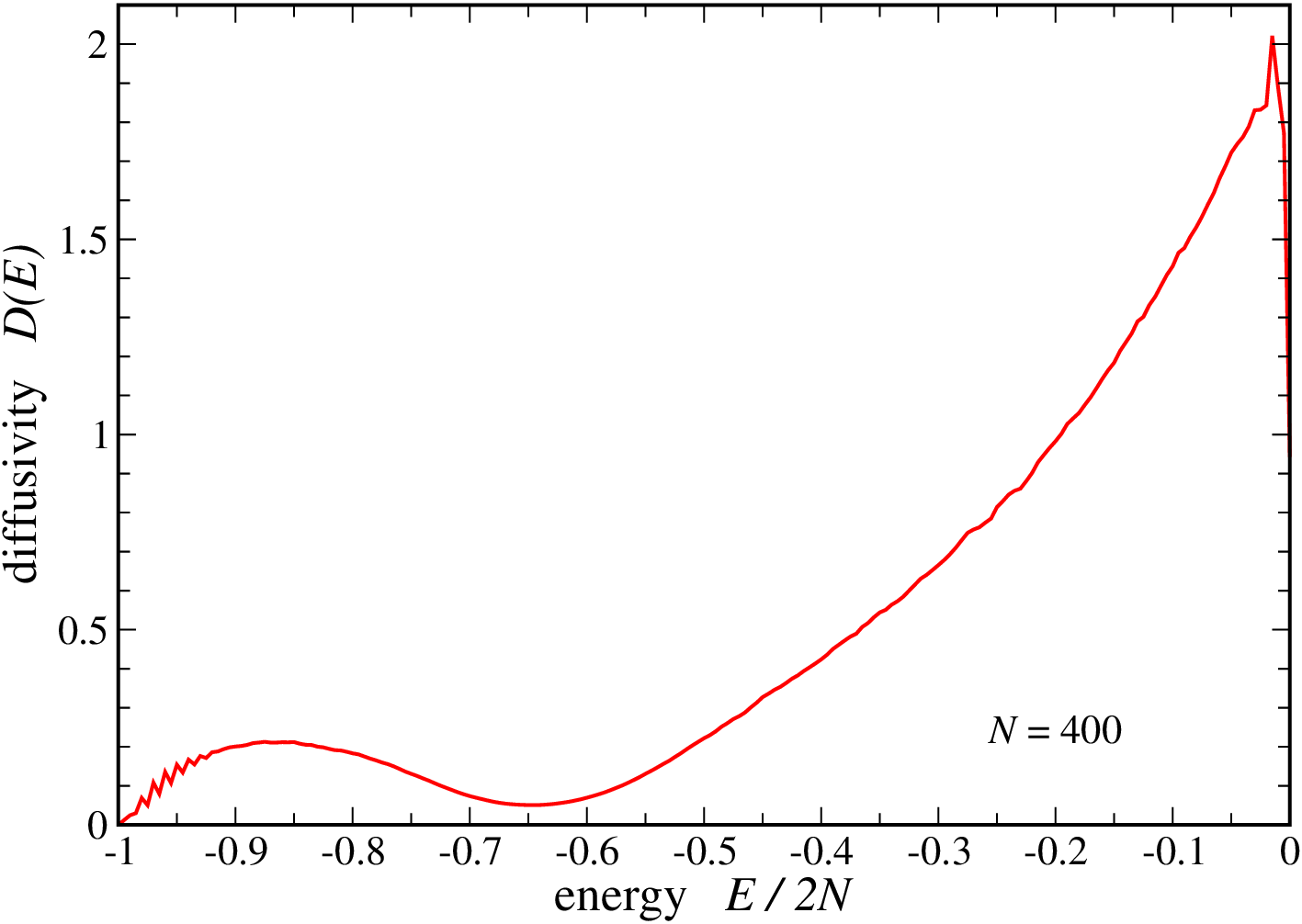}
  \end{center}
  \caption{
Local diffusivity $D(E)$ of a random walk sampling a flat histogram in energy space for the two-dimensional ferromagnetic Ising model with $N=20\times20$ spins. The local diffusivity strongly depends on the energy with a strong suppression around the critical energy $E_c \approx -1.41~N$ and the ground-state energy $E_0 = -2N$.}
  \label{fig:Diffusivity}
\end{figure}

This loss of information in the projection of the random walk in configuration space has important consequences for the random walk in energy space. Most strikingly, the local diffusivity of a random walker in energy space, which for a diffusion time $t_D$ is defined as
\begin{equation}
   D(E, t_D) = \langle (E(t) - E(t+t_D))^2 \rangle / t_D
\end{equation}
is {\em not} independent of the location in energy space. This is illustrated in Fig.~\ref{fig:Diffusivity} for the two-dimensional Ising ferromagnet.
Below the phase transition around $E_c\sim-1.41 N$ a clear minimum evolves in the local diffusivity. In this region large ordered domains are formed and by moving the domain boundaries through local spin flips only small energy changes are induced resulting in a suppressed local diffusivity in energy space. 

Because of the strong energy dependence of the local diffusivity the simulation of a multicanonical ensemble sampling all energy levels equally often turns out to be suboptimal \cite{Dayal04}. The performance of flat-histogram algorithms can be quantified for classical spin models such as the ferromagnet where the number of energy levels is given by $[-2N, +2N]$ and thereby scales with the number of spins $N$ in the system. When measuring the typical round-trip time between the two extremal energies for multicanonical simulations, these round-trip times $\tau$ are found to scale like 
\begin{equation}
  \tau \sim N^2 L^z \;,
\end{equation}
showing a power-law deviation from the $N^2$-scaling behavior of a completely unbiased random walk.
Here $L$ is the linear system size and $z$ a critical exponent describing the slow-down of a multicanonical simulation in the proximity of a phase transition \cite{Dayal04,Wu05}. The value of $z$ strongly depends on the simulated model and the dimensionality of the problem. In two dimensions the exponent increases from $z=0.74$ for the ferromagnet as one introduces competing interactions leading to frustration and disorder. The exponent becomes  $z=1.73$ for the two-dimensional fully frustrated Ising model which is defined by a Hamiltonian
\begin{equation}
H=\sum_{\langle i,j\rangle} J_{ij} \sigma_i\sigma_j \;,
\end{equation}
where the spins around any given plaquette of four spins are frustrated, e.g. by choosing the couplings along three bonds to be $J_{ij}=-1$ (ferromagnetic) and $J_{ij}=+1$ (antiferromagnetic) for the remaining bond. 
For the spin glass ,where the couplings $J_{ij}$ are randomly chosen to be +1 or -1, exponential scaling ($z=\infty$) is found \cite{Dayal04,Alder04}. 
Increasing the spatial dimension of the ferromagnet the exponent is found to decrease as $z\approx1.81,  0.74$ and $0.44$ for dimension $d=1,2$ and 3 and $z$ vanishes for the mean-field model in the limit of infinite dimensions \cite{Wu05}.

\section{Optimized ensembles}

The observed polynomial slow-down of the multicanonical ensemble poses the question whether for a given model there is an optimal choice of the histogram $H_{\rm optimal}(E)$ and corresponding weights $W_{\rm optimal}(E)$, which eliminates the slow-down. To address this question an adaptive feedback algorithm has recently been introduced that iteratively improves the weights in an extended ensemble simulation leading to further improvements in the efficiency of the algorithm by several orders of magnitude \cite{Trebst04}. The scaling for the optimized ensemble is found to scale like $O( [N \ln N]^2)$ thereby reproducing the behavior of an unbiased Markovian random walk up to a logarithmic correction. 

At the heart of the algorithm lies the idea to maximize a current $j$ of walkers that move from the lowest energy level, $E_-$, to the highest energy level, $E_+$, or vice versa, in an extended ensemble simulation by varying the weights $W_{\rm extended}(E)$. To measure the current a label is added to the walker that indicates which of the two extremal energies the walker has visited most recently.
The two extrema act as ``reflecting'' and ``absorbing" boundaries for the labeled walker: e.g., if the label is plus, a visit to $E_+$ does not change the label, so this is a ``reflecting'' boundary. However, a visit to $E_-$ does change the label, so the plus walker is absorbed at that boundary. The behavior of the labeled walker is {\em not} affected by its label except when it visits one of the extrema and the label
changes.

For the random walk in energy space, two histograms are recorded, $H_+(E)$ and $H_-(E)$, which for sufficiently long simulations converge to steady-state distributions which satisfy $H_+(E) + H_-(E) = H(E) = W(E)g(E)$. For each energy level the fraction of random walkers which have label ``plus" is then given by $f(E) = H_+(E)/H(E)$. The above-discussed boundary conditions dictate $f(E_-)=0$ and $f(E_+)=1$.

The steady-state current to first-order in the derivative is
\begin{equation}
j=D(E)H(E){{df}\over{dE}}~, \label{Eq:Current}
\end{equation}
where $D(E)$ is the walker's diffusivity at energy $E$.  There is no current if $f(E)$ is constant, since this corresponds to the equilibrium state. Therefore the current is to leading order proportional to $df/dE$. Rearranging the above equation and integrating on both sides, noting that $j$ is a constant and $f$ runs from 0 to 1, one obtains
\begin{equation}
{1\over{j}}=\int_{E_-}^{E_+}{{dE}\over{D(E)H(E)}}~.
\end{equation}
To maximize the current and thus the round-trip rate, this integral must be minimized. However, there is a constraint: $H(E)$ is a probability distribution and must remain normalized which can be enforced  with a Lagrange multiplier:
\begin{equation}
\int_{E_-}^{E_+}dE\left({1\over{D(E)H(E)}}+\lambda H(E)\right)~.
\label{Eq:Integrand}
\end{equation}
To minimize this integrand, the ensemble, that is the weights $W(E)$ and thus the histogram $H(E)$ are varied. At this point it is assumed that the dependence of $D(E)$ on the weights can be neglected. 

The optimal histogram, $H_{\rm optimal}(E)$, which minimizes the above integrand and thereby maximizes the current $j$ is then found to be
\begin{equation}
H_{\rm optimal}(E) \propto {1\over{\sqrt{D(E)}}}.
\end{equation}
Thus for the optimal ensemble, the probability distribution of sampled energy levels is simply inversely proportional to the square root of the local diffusivity.

\begin{figure}[t]
  \begin{center}
    \includegraphics[width= 0.75\textwidth]{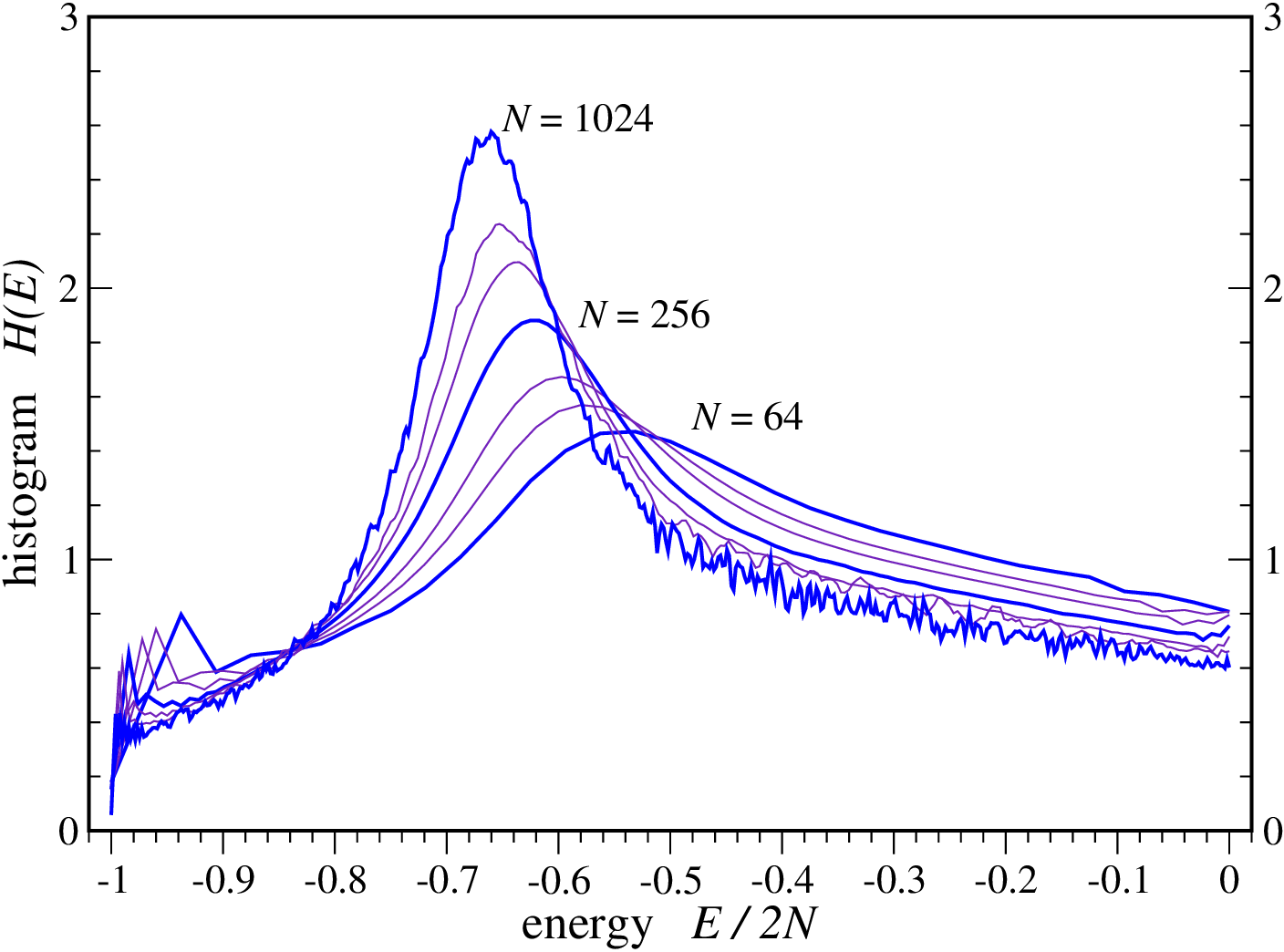}
  \end{center}
  \caption{
Optimized histograms for the two-dimensional ferromagnetic Ising model for various system sizes. After the feedback of the local diffusivity a peak evolves near the critical energy of the transition $E_c \approx -1.41~N$. The feedback thereby shifts additional resources towards the bottleneck of the simulation which are identified by a suppressed local diffusivity.}
  \label{fig:Histogram}
\end{figure}

The optimal histogram can be approximated in a feedback loop of the form

\begin{itemize}
\item Start with some trial weights $W(E)$, e.g. $W(E)=1/g(E)$.
\item Repeat
\begin{itemize}
\item Reset the histograms $H(E) = H_+(E) = H_-(E) = 0$.
\item Simulate the system with the current weights for $N$ sweeps:
\begin{itemize}
\item Updates are accepted with probablity $\min[1, W(E')/W(E)]$.
\item Record the histograms $H_+(E)$ and $H_-(E)$.
\end{itemize}
\item From the recorded histogram an estimate of the local diffusivity is obtained as 
\vspace{2.5mm} \[
D(E) \propto {{1}\over{H(E){{df}\over{dE}}}}\;, 
\] \vspace{2.5mm}
where the fraction $f(E)$ is given by
$f(E) = H_+(E)/H(E)$
and $H(E)$ is the histogram $H(E) = H_+(E) + H_-(E)$.
\item Define new weights as
\vspace{2.5mm} \[
   W_{\rm optimized}(E) = W(E) \sqrt{\frac{1}{H(E)} \cdot \frac{df}{dE} } \;.
\] \vspace{2.5mm}
\item Increase the number of sweeps for the next iteration \\ $N_{\rm sweeps} \leftarrow 2N_{\rm sweeps}$.
\end{itemize}
\item Stop once the histogram $H(E)$ has converged.
\end{itemize}

The implementation of this feedback algorithm requires to change only a few lines of code in the original local-update algorithm for the Ising model. Some additional remarks are useful:  
\begin{enumerate}
\item In contrast to the Wang-Landau algorithm, the weights $W(E)$ are modified only after a batch of $N_{\rm sweeps}$ sweeps, thereby ensuring detailed balance between successive moves at all times.
\item The initial value of sweeps $N_{\rm sweeps}$ should be chosen large enough that a couple of round trips are recorded, thereby ensuring that steady state data for $H_+(E)$ and $H_-(E)$ are measured.
\item The derivative $df/dE$ can be determined by a linear regression, where the number of regression points is flexible. Initial batches with the limited statistics of only a few round trips may require a larger number of regression points than subsequent batches with smaller round-trip times and better statistics. 
\item Similar to the multicanonical ensemble, the weights $W(E)$ can become very large and storing the logaritms may be advantageous. The reweighting then becomes $\ln W_{\rm optimized}(E) = \ln W(E) + [\ln \frac{df}{dE} -\ln {H(E)} ]/2$~.
\end{enumerate}

At the end of the simulation, the density of states can be estimated from the recorded histogram as $g(E) = H_{\rm optimized}(E)/W_{\rm optimized}(E)$ and normalized as described above.

Fig.~\ref{fig:Histogram} shows the optimized histogram for the two-dimensional ferromagnetic Ising model. The optimized histogram is no longer flat, but a peak evolves at the critical region around $E_c \approx -1.41~N$ of the transition. The feedback of the local diffusivity reallocates resources towards the bottlenecks of the simulation which have been identified by a suppressed local diffusivity.

\begin{figure}[t]
  \begin{center}
    \includegraphics[width= 0.75\textwidth]{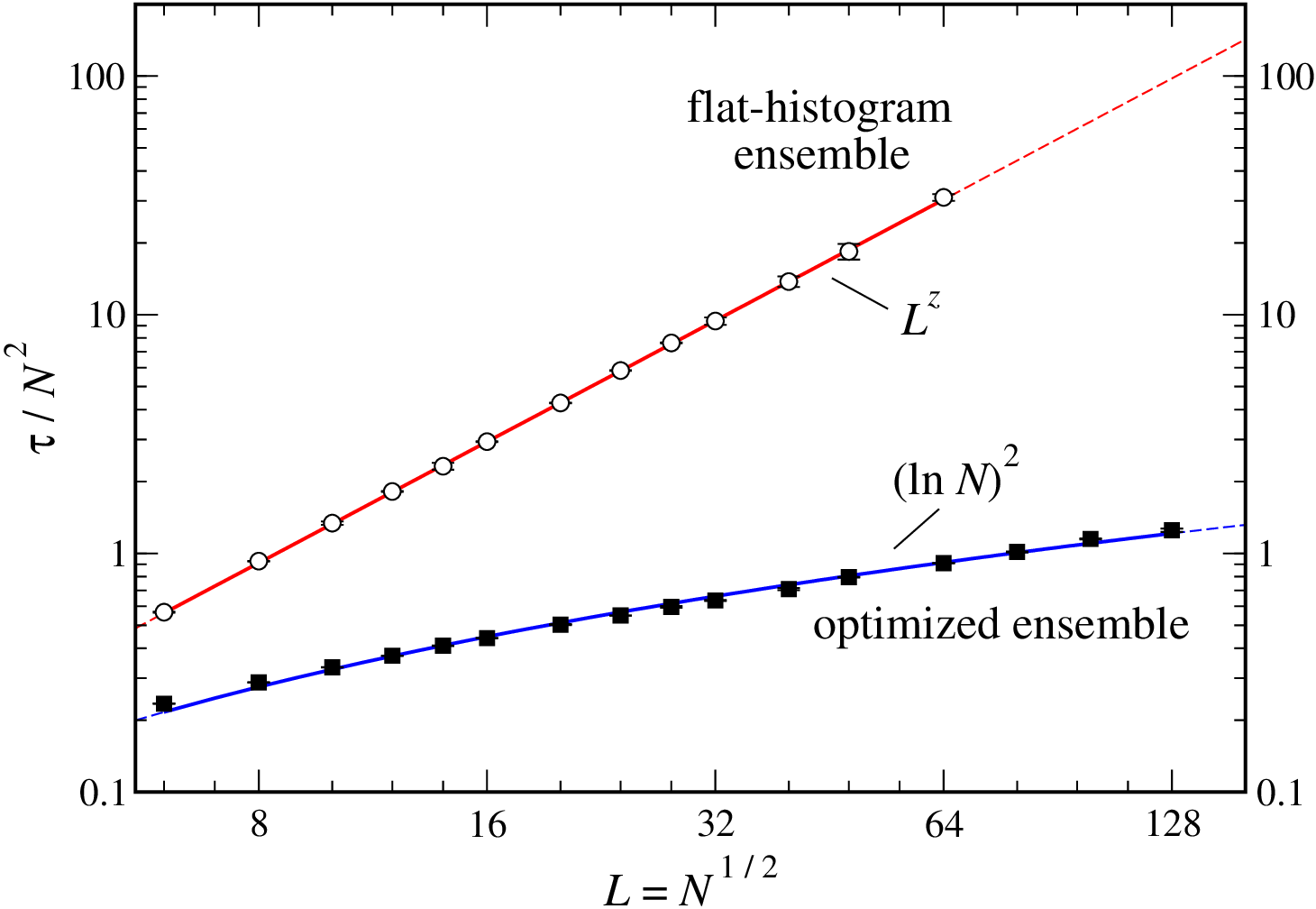}
  \end{center}
  \caption{
Scaling of round-trip times for a random walk in energy space sampling a flat histogram (open squares) and the optimized histogram (solid circles) for the two-dimensional fully frustrated Ising model. While for the multicanonical simulation a power-law slow-down of the round-trip times $O(N^2L^z)$ is observed, the round-trip times for the optimized ensemble scale like $O( [N \ln N]^2)$ thereby approaching the ideal $O(N^2)$-scaling of an unbiased Markovian random walk up to a logarithmic correction.}
  \label{fig:Scaling}
\end{figure}

When analyzing the scaling of round-trip times for the optimized ensemble one finds a considerable speedup: The power-law slow-down of round-trip times for the flat-histogram ensemble $O(N^2L^z)$ is reduced to $O( [N \ln N]^2)$ for the optimized ensemble, e.g. there is only a logarithmic correction to the scaling of a completely unbiased random walk with $O(N^2)$-scaling. 
For the two-dimensional fully frustrated Ising model the scaling of round-trip times is shown in Fig.~\ref{fig:Scaling}. 
This scaling improvement results in a speedup by a nearly two orders of magnitude already for a system with some $128 \times 128$ spins.

\section{Simulation of dense fluids}

Extended ensembles cannot only be defined as a function of energy, but in arbitrary reaction coordinates $\vec{R}$ onto which a random walk in configuration space can be projected. The generalized weights in these reaction coordinates $W_{\rm extended}(\vec{R})$ are then used to bias the random walk along the reaction coordinate by accepting moves from a configuration $c_1$ with reaction coordinate $\vec{R}_1$ to a configuration $c_2$ with reaction coordinate  $\vec{R}_2$ with probability
\begin{equation}
 p_{\rm acc} (c_1 \rightarrow c_2 ) = p_{\rm acc} (\vec{R}_1 \rightarrow \vec{R}_2 )
  = \min \left( 1, \frac{W_{\rm extended}(\vec{R}_2)}{W_{\rm extended}(\vec{R}_1)} \right) \;.
\end{equation}
The generalized weights $W_{\rm extended}(\vec{R})$ can again be chosen in such a way that similar to a multicanonical simulation a flat histogram is sampled along the reaction coordinate by setting the weights to be inversely proportional to the density of states defined in the reaction coordinates, that is $W_{\rm extended}(\vec{R}) \propto 1 / g(\vec{R})$. 

An optimal choice of weights can be found by measuring the local diffusivity of a random walk along the reaction coordinates and by applying the feedback method to shift weight towards the bottlenecks in the simulation. This generalized ensemble optimization approach has recently been illustrated for the simulation of dense Lennard-Jones fluids close to the vapor-liquid equilibrium \cite{Trebst05}.
The interaction between particles in the fluid is described by a pairwise Lennard-Jones potential of the form
\begin{equation}
   \Phi_{\rm{LJ}}(R) = 4\epsilon 
      \left[ \left(\frac{\sigma}{R}\right)^{12} -  \left(\frac{\sigma}{R}\right)^6 \right] \;,
   \label{Eq:LJ}
\end{equation}
where $\epsilon$ is the interaction strength, $\sigma$ a length parameter, and $R$ the distance between two particles. It is this distance $R$ between two arbitrarily chosen particles in the fluid that one can use as a new reaction coordinate for a projected random walk. For a given temperature defining an extended ensemble with weights $W_{\rm extended}(R)$ and recording a histogram $H(R)$ during a simulation will then allow to calculate the pair distribution function $g(R) = H(R) / W_{\rm extended}(R)$.
The pair distribution function $g(R)$ is closely related to the potential of mean force (PMF)
\begin{equation}
  \Phi_{\rm{PMF}}(R) = - \frac{1}{\beta} \ln g(R) \;,
\end{equation}
which describes the average interaction between two particles in the fluid in the presence of many surrounding particles.

\begin{figure}[t]
  \begin{center}
    \includegraphics[width= 0.75\textwidth]{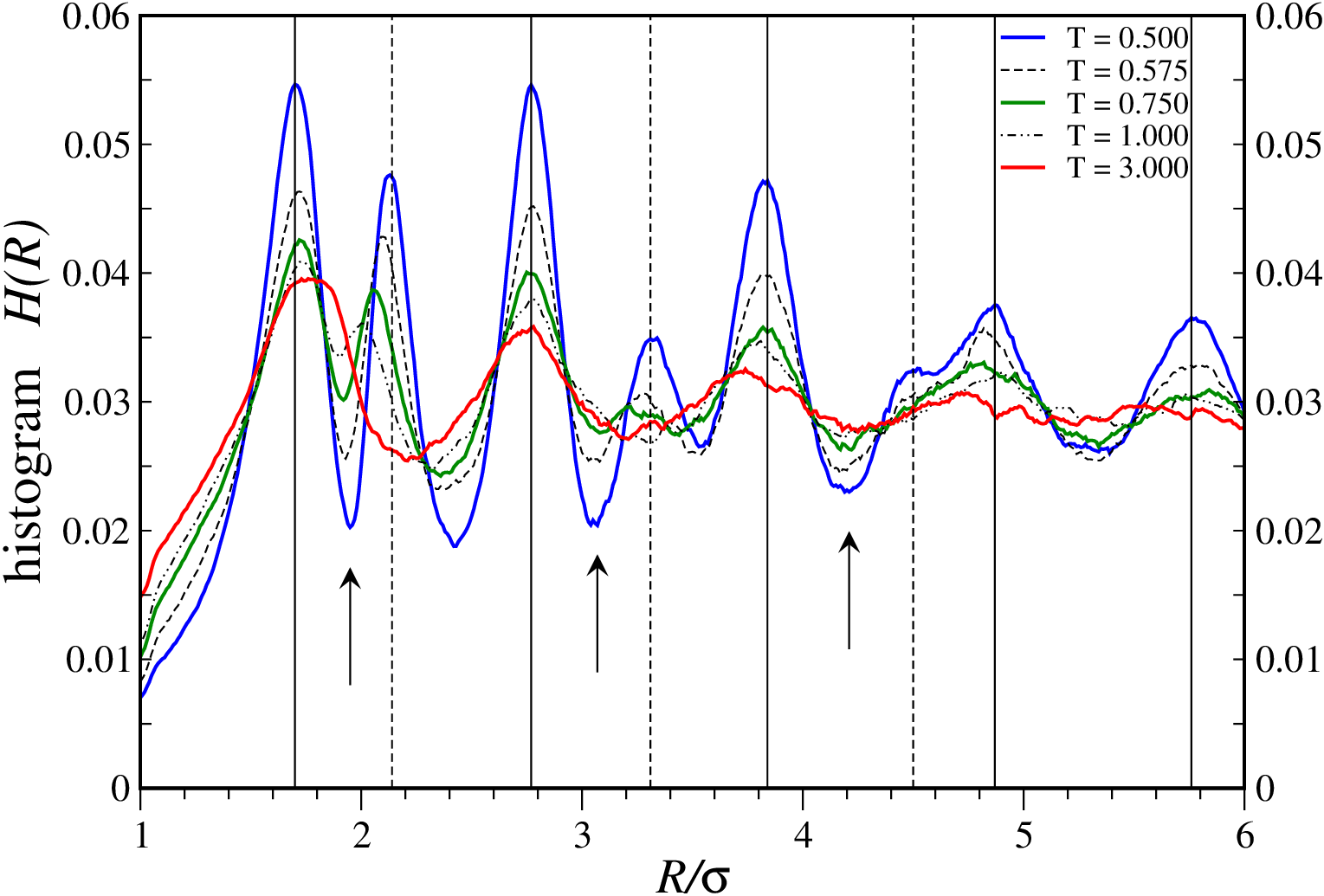}
  \end{center}
  \caption{Optimized histograms for a dense two-dimensional Lennard-Jones fluid after feedback of the 
  local diffusivity in a radial coordinate for varying temperatures. 
  For the optimized ensemble peaks evolve at the free energy barriers between the shells of the liquid 
  which proliferate at lower temperatures (solid lines). 
  Additional peaks emerge at low temperatures (dashed lines) revealing interstitial states (arrows) 
  between the shells of the liquid.}
  \label{fig:LJ-Histograms}
\end{figure}

For high particle densities and low enough temperatures shell structures form in the fluid which are reminiscent of the hexagonal lattice of the solid structure at very low temperatures. These shell structures are revealed by a sinusoidal modulation in the PMF. Thermal equilibration between the shells is suppressed by entropic barriers which form between the shells. Again, one can ask what probability distribution, or histogram, should be sampled along the reaction coordinate, in this case the radial distance $R$, so that equilibration between the shells is improved. Measuring the local diffusivity for a random walk along the radial distance $R$ in an interval $[R_{\rm min}, R_{\rm max}]$ and subsequently applying the feedback algorithm described above optimized histograms $H(R)$ are found which are plotted in Fig.~\ref{fig:LJ-Histograms} for varying temperatures \cite{Trebst05}. 
The feedback algorithm again shifts additional weight in the histogram towards the bottleneck of the simulation, in this case towards the barriers between the shells. Interestingly, additional peaks emerge in the optimized histogram as the temperature is lowered towards the vapor-liquid equilibrium. The minima between these peaks point to additional meta-stable configurations which occur at these low temperatures, namely interstitial states which occur as the shells around two particles merge as detailed in Ref.~\cite{Trebst05}. 

This example illustrates that for some simulations the local diffusivity and optimized histogram {\em themselves} are very sensitive measures that can reveal interesting underlying physical phenomena which are otherwise hard to detect in a numerical simulation. In general, a strong modulation of the local diffusivity for the random walk along a given reaction coordinate is a good indicator that the reaction coordinate itself is a good choice that captures some interesting physics of the problem.

\section{Parallel tempering / replica-exchange methods}
\label{sec:pt}
The simulation of frustrated and/or disordered systems suffers from a similar tunneling problem than the simulation of first-order phase transitions: local minima in energy space are separated by barriers that grow with system size. While the multicanonical or optimized ensembles do not help with the NP-hard problems faced by spin glasses, they are efficient in speeding up simulations of frustrated magnets without disorder \cite{Trebst04}.

An alternative to these extended ensembles for the simulation of frustrated magnets is the ``parallel tempering" or ``replica-exchange" Monte Carlo method \cite{Swendsen86,Marinari92,Lyubartsev92,HukushimaNemoto96}. Instead of performing a single simulation at a fixed temperature, simulations are performed for $M$ replicas at a set of temperatures $T_1, T_2, \ldots,T_M$. In addition to standard Monte Carlo updates at a fixed temperature, exchange moves are proposed to swap two replicas between adjacent temperatures. These swaps are accepted with a probability 
\begin{equation}
\label{eq:swaprate}
\min[1,\exp(\Delta\beta\Delta E)],
\end{equation}
where $\Delta\beta = \beta_j - \beta_i$ is the difference in inverse temperatures and $\Delta E = E_j-E_i$ the difference in energy between the two replicas $i$ and $j$. 

The effect of these exchange moves is that a replica can drift from a local free energy minimum at low temperatures to higher temperatures, where it is easier to cross energy barriers and equilibration is fast. Upon cooling (by another sequence of exchanges) it can end up in a different local minimum on time scales that are much shorter compared to a single simulation at a fixed low temperature. 
This random walk of a single replica in temperature space is the conjugate analog of the random walk in energy space discussed for the extended ensemble techniques. The complement of the statistical ensemble, defined by the weights $W_{\rm extended}(E)$, is the particular choice of temperature points in the temperature set $\{T_1, T_2, \ldots, T_M\}$ for the parallel tempering simulation. The probability of sampling any given temperature $T$ in an interval $T_i < T < T_{i+1}$ can then be approximated by ${H}(T) \propto 1/\Delta T$, where $\Delta T = T_{i+1} - T_i$ is the length of the temperature interval around the temperature $T$. This probability distribution ${H}(T)$ is the equivalent to the histogram $H(E)$ in the extended ensemble simulations. The ensemble optimization technique discussed above can thus be reformulated to optimize the temperature set in a parallel-tempering simulation in such a way that the rate of round trips between the two extremal temperatures, $T_1$ and $T_M$ respectively, is maximized \cite{Katzgraber06,Trebst06}.

\begin{figure}[t]
  \begin{center}
    \includegraphics[width= 0.75\textwidth]{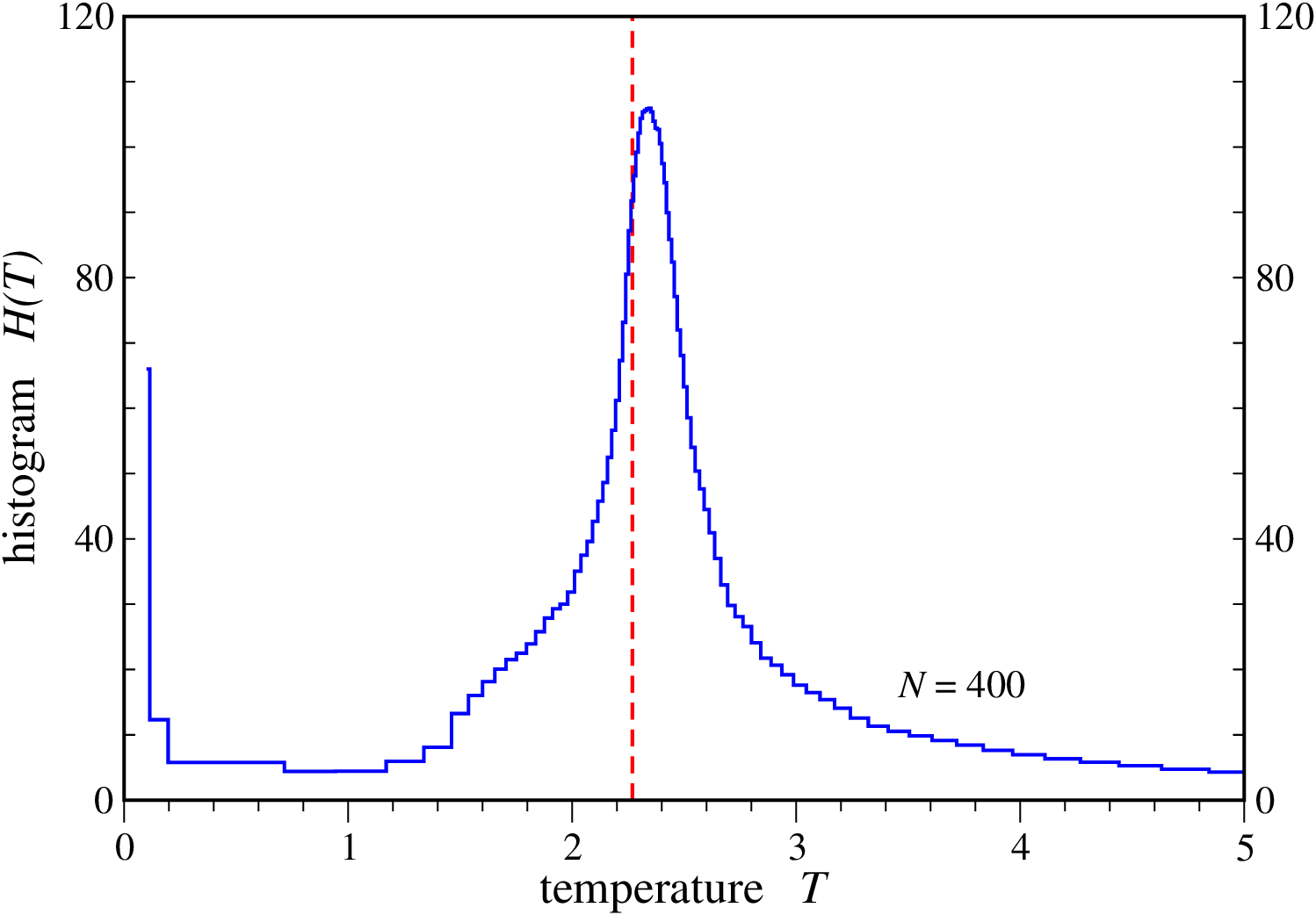}
  \end{center}
  \caption{
Optimized temperature distribution $H(T)$ for a two-dimensional Ising ferromagnet with $N=20\times20$ spins and 100 replicas/temperature points. After the feedback of the local diffusivity a peak evolves near the critical temperature of the transition $T_c \approx 2.269$. The feedback thereby shifts additional resources towards the bottleneck of the simulation which are identified by a suppressed local diffusivity. Note the similarity to Fig.~\ref{fig:Histogram}.}
  \label{fig:THistogram}
\end{figure}

\begin{figure}[t]
  \begin{center}
    \includegraphics[width= 0.75\textwidth]{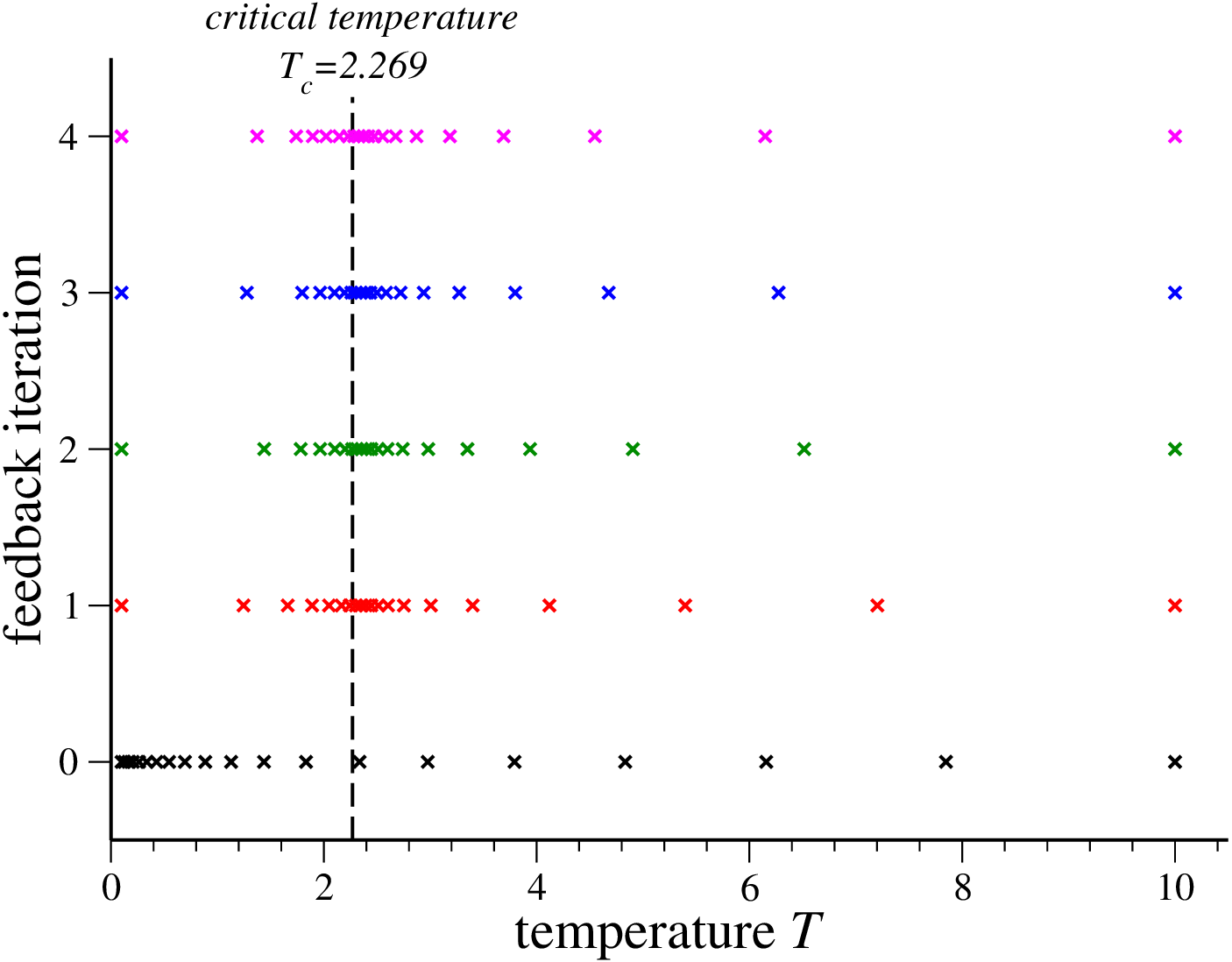}
  \end{center}
  \caption{
Optimized temperature sets for a two-dimensional Ising ferromagnet with $N=20\times20$ spins. The initial temperature set with 20 temperature points is determined by a geometric progression \cite{Katzgraber06,Predescu04} for the temperature interval $[0.1, 10]$. After feedback of the local diffusivity the temperature points accumulate near the critical temperature $T_c=2.269$ of the phase transition (dashed line). Similar to the ensemble optimization in energy space the feedback of the local diffusivity relocates resources towards the bottleneck of the simulation. }
  \label{fig:TSets}
\end{figure}

Starting with an initial temperature set $\{T_1, T_2, \ldots, T_M\}$ a parallel tempering simulation is performed where each replica is labeled either ``plus" or ``minus" indicating which of the two extremal temperatures the respective replica has visited most recently. This allows to measure a current of replicas diffusing from the highest to the lowest temperature by recording two histograms, $h_+(T)$ and $h_-(T)$ for each temperature point. The current $j$ is then given by
\begin{equation}
  j = D(T) {H}(T) \frac{df}{dT} \;,
  \label{eq:Tcurrent}
\end{equation}
where $D(T)$ is the local diffusivity for the random walk in temperature space, and $f(T) = h_+(T)/[h_+(T) + h_-(T)]$ is the fraction of random walkers which have visited the highest temperature $T_M$ most recently. The probability distribution ${H}(T)$ is normalized, that is
\begin{equation}
  \int_{T_1}^{T_M} {H}(T)~ dT = C \int_{T_1}^{T_M}  \frac{dT}{\Delta T}  = 1 \;,
\end{equation}
where $C$ is a normalization constant. Rearranging equation (\ref{eq:Tcurrent}) the local diffusivity $D(T)$ of the random walk in temperature space can be estimated as
\begin{equation}
  D(T) \propto \frac{\Delta T}{df/dT} \; .
\end{equation}
In analogy to the argument for the extended ensemble in energy space the current $j$ is maximized by choosing a probability distribution
\begin{equation}
  {H}_{\rm optimal}(T) \propto \frac{1}{\sqrt{D(T)}} \propto \sqrt{\frac{1}{\Delta T}~ \frac{df}{dT}}
\end{equation}
which is inversely proportional to the square root of the local diffusivity.
The optimized temperature set $\{T'_1, T'_2, \ldots, T'_M\}$ is then found by choosing the $n$-th temperature point $T'_n$ such that
\begin{equation}
   \int_{T'_1}^{T'_n} {H}_{\rm optimal}(T)~ dT = \frac{n}{M} \;,
\end{equation}
where $M$ is the number of temperature points in the original temperature set, and the two extremal temperatures $T_1' = T_1$ and $T_M'=T_M$ remain unchanged. Similarly to the algorithm for the ensemble optimization this feedback of the local diffusivity should be iterated until the temperature set is converged.

Figures \ref{fig:THistogram} and \ref{fig:TSets} illustrate the so-optimized temperature sets for the Ising ferromagnet obtained by several iterations of the above feedback loop. After the feedback of the local diffusivity, temperature points accumulate near the critical temperature $T_c=2.269$ of the transition. This is in full analogy to the optimized histograms for the extended ensemble simulations where resources are shifted towards the critical energy of the transition, for comparison see Figs.~\ref{fig:Histogram} and \ref{fig:THistogram}. 

It is interesting to note that for the so-optimized temperature set the acceptance rates for swap moves are not independent of the temperature \cite{Katzgraber06}. Around the critical temperature, where temperature points are accumulated by the feedback algorithm, the acceptance rates are higher than at higher/lower temperatures, where the density of temperature points becomes considerably smaller after feedback \cite{Katzgraber06,Trebst06}. The almost Markovian scaling behavior for the optimized random walks in either energy or temperature space is thus generated by a problem-specific statistical ensemble which is characterized neither  by a flat histogram nor flat acceptance rates for exchange moves, but by a characteristic probability distribution which concentrates resources at the minima of the measured local diffusivity.

\section{Outlook}

The ensemble optimization technique reviewed in this chapter should be broadly applicable to a wide range of applications -- possibly speeding up existing uniform sampling techniques by orders of magnitude. It has recently been used to improve broad-histogram Monte Carlo techniques \cite{Trebst04} as well as parallel-tempering Monte Carlo simulations \cite{Katzgraber06}, with applications to frustrated and disordered spin systems \cite{Trebst04,Katzgraber06}, dense fluids \cite{Trebst05}, as well as folded proteins \cite{Trebst06}. It also holds promise to improve the simulation of quantum systems close to a phase transition when optimizing the extended ensemble introduced for the quantum Wang-Landau algorithm outlined in Ref.~\cite{QWL}.

\end{document}